# The Impact of Phosphate Fertilizer Industry Consolidation on Future Phosphorus Supply for World Agriculture

**Authors**: Anna Shchiptsova[1,2], Michael Obersteiner[1,3]

1: International Institute for Applied Systems Analysis, Schlossplatz 1, A-2361 Laxenburg, Austria

2: HSE University, St. Petersburg 190121, Russia

2: Environmental Change Institute, University of Oxford, South Parks Road, Oxford, OX1 3QY, United Kingdom

**Abstract.** The addition of phosphorus, in the form of mineral fertilizer, becomes necessary in most agricultural soils in order to achieve consistent high yield levels of intensive farming and maintain soil fertility. Recent consolidation of phosphate fertilizer industry has transformed fragmented trade into a single integrated global network, where a small group of large-scale companies dominates the international market for phosphate commodity fertilizers. To assess the impact of new trade structure on future region-level phosphorus supply, we simulate behavior of markets for ammonium phosphates in the FAO scenarios of global intensive farming evolution. Details of market microstructure are represented here by a many-to-many matching market. Current spatial distribution of global demand in ammonium phosphates is projected to strengthen by 2030. Bootstrap simulations produce similar network structures for both scenarios showing reduction in the density of the distributed market. In response to the non-uniform demand growth across regions, market concentration is expected to increase for small-scale markets, and to remain predominantly stable for large-scale markets; on the supply side, simulated equilibria point out large-scale multi-market suppliers concentrating on fewer markets than before. A high rate of import substitution by local suppliers in some markets indicate the need of additional region-level capital investment.

## 1 Introduction

The addition of phosphorus, in the form of mineral fertilizer, becomes necessary in most agricultural soils in order to achieve consistent high yield levels of intensive farming and maintain soil fertility (McLaughlin et al. 2011). Industrial fertilizer production consumes 90% of phosphorus (Campbell et al. 2017) mined at commercial levels from naturally occurring deposits of phosphate-bearing rocks. At the field, phosphorus fertilization comprises building up soil phosphorus reserves and replacing phosphorus that has been removed with the crop at harvest to ensure optimal farm output and efficient use of all nutrient inputs (Reetz 2016). With logistical constraints on manure application (Syers et al. 2008) and currently undeveloped large-scale cost-competitive technologies of phosphorus recovery and reuse (Mayer et al. 2016), the fertilizer industry is set to grow in the next decade (Gilbert 2009; IFA 2018) with increasing global demand for agricultural crops (Tilman et al. 2011); spatial patterns in phosphorus demand will continue to shift with changes in cropping (Johnston and Dowson 2005) driven by sustainable and economically workable agricultural intensification (Pretty and Bharucha 2014).

The process of phosphate fertilizer production encompasses large-scale operations of mining and chemical processing (Harben 2002). Manufacturers gain competitive edge in the business through access to cheap and abundant feedstock, access to low-cost credit and low processing and distribution costs created through economies of scale and value chain integration (IFA 2018). The structure of the supply chain and geographical concentration of phosphorus reserve base have pushed the industry towards consolidation. The number of phosphate producers in the US dropped from 18 firms in 1990 to 12 firms in 2000, and to 4 firms in 2018 (Mack 2015; Nutrien 2019) after a merger of IMC Global and Cargill's crop nutrition division into the Mosaic Company in 2004, Mosaic's acquisition of phosphate operations of CF Industries in 2014, and a merger of PotashCorp and Agrium in 2018. Since the price spike in 2008, the increased price volatility of phosphate products has strengthened positions of companies with highly integrated business models (McGill 2014). Having an exclusive access to 73% of world phosphate deposits, OCP Group (Morocco) adopted an investment strategy to double its rock extraction capacity and triple its phosphate processing capacity in the period from 2008 to 2027 (OCP 2017). To diversify oil-centered economy, Saudi Arabia invested into fully integrated phosphate projects utilizing national resources of phosphate, natural gas, and sulfur (Aldagheiri 2016). Saudi Arabian Mining Company (Ma'aden) began operations at the world's largest phosphate fertilizer complex in 2012, doubled company's production capacity in 2018, and planned construction of the third large-scale fertilizer complex by 2025 (Ma'aden 2019). The manufacturing assets in the Former Soviet Union were consolidated by PhosAgro and EuroChem business groups (both established in 2001), which acquired fine quality phosphate mines in Northwest Russia and adapted strongly integrated operational structure including ammonia, sulfuric acid and natural gas production (PhosAgro 2019; EuroChem 2019). Being the number one consumer of phosphates, China is largely self-sufficient in phosphorus products (Fabbe et al. 2018); the structure of Chinese market is opaque, though it is known that there are three major players (GPCG, Yihua and YTH) alongside many smaller ones (McGill 2014). China is in the process of consolidating its domestic industry (McGill 2014). In 2019, Wengfu and Kailin companies merged into GPCG to form the third largest phosphate fertilizer producer behind OCP and Mosaic (PhosAgro 2020). Today the phosphate fertilizer industry exhibits high entry barriers to new large-scale phosphate projects due to the combined factors of cheap feedstock access and long period required to recoup the capital invested (IFA 2018).

Commodity fertilizer sales generate a major part of phosphate company's revenue (Fabbe et al. 2018; IFA 2018). In 2017, 69% of phosphate fertilizers (IFA 2020a, IFA 2020b) were consumed in the form of ammonium phosphates (DAP and MAP), which can be applied directly to soil or used as a raw material for blends (Harben 2002). Since 2008, the price of ammonium phosphates has become a benchmark price for other phosphorus products (McGill 2014). DAP/MAP fertilizer trade is global, with customer preferences being largely determined by the product price (Mosaic 2019). Following consolidation on production side of the industry, the markets of these bulk commodities have transformed into a single integrated global network. In 2017, a small group of large-scale businesses (Mosaic, OCP, Ma'aden,

PhospAgro, EuroChem and China's export) was responsible for 39% of DAP/MAP sales (UN Comtrade 2020; companies data); the remaining trade was fragmented with domestic producers and smaller exporters capturing a part of regional demand through geographical proximity and, as a result, lower transportation costs (McGIll 2014). In this paper, we take a step towards a quantitative impact assessment of phosphate fertilizer industry consolidation on the markets for ammonium phosphates. We utilize the existing information about the underlying market microstructure to account for significant factors influencing DAP/MAP price dynamics. The open data on annual industrial production and trade in ammonium phosphates, as well as the country- and crop-level data on phosphorus fertilizer use in agriculture, are integrated into a many-to-many matching model of spatially distributed DAP/MAP commodity market. The model operates under the assumption that commodity fertilizers prevail in the international phosphorus trade during the next decade, which corresponds to phosphate fertilizer industry development in the IFA "Commodity Classic" scenario for 2030 (IFA 2018). The competitive equilibrium in the two-sided matching market is used here to assess spatial patterns of phosphorus demand in different global crops production scenarios (FAO 2018) regarding their impact on the amount of competition in the integrated trade network, and, as a result, their impact on the structure of future phosphorus supply for world agriculture. The scenarios investigate global intensive farming evolution by year 2030 under current or increasing economic and technological inequalities between countries.

The rest of the paper is organized as follows. In Section 2, we describe market microstructure and price dynamics, and present the matching model of spatially distributed DAP/MAP commodity market. Section 3 contains detailed information about the model application to global crops production scenarios. The results of computational experiments are summarized in Section 4 and put into perspective through discussion in Section 5. Some final remarks are given in Section 6.

## 2 Methodology

*2.1 Market microstructure*

World trade in DAP/MAP commodities breaks down into transactions on individual markets residing at different geographical locations. On the supply side, there exists many one-market suppliers within market vicinity, and a limited number of large-scale multi-market international sellers. International suppliers differentiate price offers for distinct markets based on bulk transportation costs and customs duties (McGill 2014). Each market has diverse customer base spanning from large-scale distributors, traders, fertilizer blenders and complex fertilizer producers to small- and medium-scale distributors and farmers, who can make direct international purchases due to container shipping (EuroChem 2019; OCP 2017). Trade is to a great extent organized by short-term bilateral contracts with business intelligence companies continuously gathering information about market transactions and publishing price point assessments for the international trade network (McGill 2014). Under market transparency, sellers and buyers make contract pricing decisions based on perceived market trends and potential offers from competing suppliers (Mew et al. 2018).

Phosphate products are subject to seasonal demand due to the timing of fertilizer field application (PhosAgro 2019). Making fertilizer available to farmers in a short time window requires pre-seasonal inventory accumulation on a market (Mosaic 2019). Contract transactions in Europe and the US peak in the first quarter of the year with Russia having the second peak in August before the winter wheat growing season. Trade in South America typically raises in the second and third quarters, while shipments to India intensify in the third quarter ahead of the rice and winter wheat planting. There remain periods of low contract activity on the markets such as in May and November (PhosAgro 2013).

Technology and economics command large-scale suppliers to operate fertilizer manufacturing facilities continuously year-round to attain high output volume with low unit costs of production. To balance practice of continuous production with demand fluctuations, producers have adopted flexible sales strategy by distributing products to all world regions and, as a result, spreading demand throughout the year (PhosAgro 2019; OCP 2017). Globally diversified sales allow international suppliers to mitigate the risk of low fertilizer demand on an individual market and sign contracts which provide the best opportunity for profit maximization in the trade network (PhosAgro 2019).

*2.2 Model description*

We now describe the model of annual DAP/MAP commodity trade in the spatially distributed global market (Figure 1) formally. In this model, a global market for the same good consists of $n$ markets residing at different locations. The markets are linked by the shared access to a finite pool of $m$ international suppliers. Let $i = 1, \ldots, m$ index suppliers and $j = 1, \ldots, n$ index markets. Suppose that each supplier $i$ has an annual production capacity of at most $s_i$ goods. Customers on market $j$ want to purchase a total of $d_j$ products. We call $x^j = (x_1^j, \ldots, x_m^j)$ annual flows of goods that international suppliers produce for market $j$. Each $s_i$ and $d_j$ is a positive whole number and each $x_i^j$ is a nonnegative whole number.

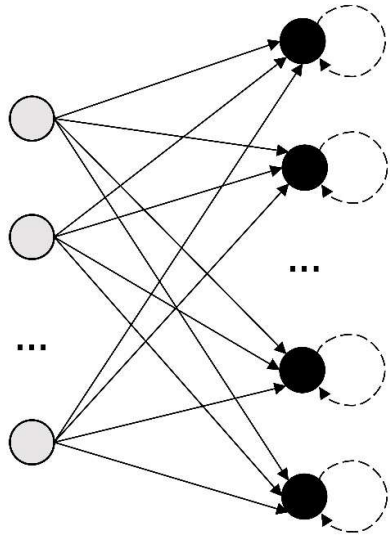

**Figure 1.** A structure of the spatially distributed global market for the same commodity. Each black node is a separate geographic commodity market; each gray node denotes an international supplier. Solid links represent trade flows from international suppliers to markets; domestic trade flows are indicated by dashed lines.

We call $c_{oj}$ the constant unit cost of production for any local supplier on market $j$. Let $a$ be a constant cost of adding one more unit of locally produced good to the market inventory. Local suppliers have an incentive to produce goods if the market price is above their incurred costs.

So given a cost structure, function $e_{oj}$ of annual customers spending on $z$ goods from local suppliers is defined to be $e_{oj}(z) = z \cdot (az + c_{oj})$. Suppose that customers on market $j$ will have to pay the constant trade cost $t_{ij}$ per unit of goods purchased from international supplier $i$. We assume that $t_{ij}$ includes production, transportation and customs costs. Let function $e_{ij}$ of annual customers spending on delivery and storage of $z$ goods at their cost value be given by $e_{ij}(z) = z \cdot (az + t_{ij})$. Then, valuation $v_j$ that customers have for $x^j$ imported goods is determined by the maximum savings they receive from substituting local goods with shipped imports (at cost value). Formally,

$$v_j(x^j) = \max\left\{e_{oj}(d_j) - e_{oj}\left(d_j - \sum_{i=1}^{m} z_i\right) - \sum_{i=1}^{m} e_{ij}(z_i) \colon z_i \in \{0,1,\dots,x_i^j\} \text{ for all } i = 1,\dots,m\right\}.$$

Suppose that each international supplier $i$ offers to sell goods with a markup $p_i \geq 0$. Then buyers on market $j$ will have to pay price $ax_i^j + t_{ij} + p_i$ for the purchase of $x_i^j$ goods from supplier $i$. If market $j$ imports $x^j$ goods at the given markups, then the customers' payoff is their valuation for these goods, minus the sum of the markups for imports: $v_j(x^j) - \sum_{i=1}^{m} p_i x_i^j$. We call an equilibrium the vector of markups $(p_1, \dots, p_m)$ and the bundle of flows $(x^1, \dots, x^n)$ that satisfy the following properties:

1. *Capacity constraints.* $\sum_{j=1}^{n} x_i^j \leq s_i$ for all $i = 1, \dots, m$.

2. *Utility maximization.* Customers want to maximize their payoff:

$x^j \in \text{argmax}\{v_j(z) - \sum_{i=1}^{m} p_i z_i \colon z_i \in \{0,1,\dots,s_i\} \text{ for all } i = 1,\dots,m\}$ for all $j = 1,\dots,n$.

3. *Market clearance.* Supplier $i$ goods are unsold ($\sum_{j=1}^{n} x_i^j = 0$) only if $p_i$ is 0 ($i = 1,\dots,m$).

A problem of finding competitive equilibrium satisfying conditions 1-3 is a special case of the many-to-many matching problem studied by Gul and Stacchetti (2000). They provide a generalization of the English auction procedure to the case in which multiple objects (imported goods) are to be sold simultaneously, and each agent (aggregate market demand) may wish to consume a bundle of different objects. The English auction stops at the equilibrium with the smallest markup vector among all equilibria satisfying conditions 1-3. This property of the solution assumes that an international supplier compensates excess customer demand for its products by markup increase.

Demand-side competition in the model is measured by the Herfindahl–Hirschman index of market concentration (Hirschman 1964). The normalized index $H_j$ of concentration for market $j$ is given by

$$H_j = \frac{\sum_{i=0}^{m}\left(\frac{x_i^j}{d_j}\right)^2 - \frac{1}{m+1}}{1 - \frac{1}{m+1}},$$

where $x_o^j$ is an amount of local goods sold on the market. This index of concentration takes the value 0 when suppliers have equal market shares and approaches 1 in the case of monopoly. The index of diversification (Berry 1971), for the supply side, is concerned with the number of markets in which international supplier is active. The normalized index $D_i$ of diversification for supplier $i$ is given by

$$D_i = 1 - \frac{\sum_{j=1}^{n}\left(\frac{x_i^j}{s_i}\right)^2 - \frac{1}{n}}{1 - \frac{1}{n}}.$$

The index of diversification takes the value 0 when an international supplier is active on a single market and attains 1 when the supplier in question sells goods in equal shares to all $n$ markets. These measures of market density and market share of local suppliers are used here to assess the structure of DAP/MAP supply for regional agricultural production.

**3 Empirical application**

*3.1 Concept*

The model is applied to the closed trade network, which accounts for all DAP/MAP trade. World demand is partitioned into regions ($n = 9$): Africa, East Asia, Eastern Europe and Central Asia, Latin America, North America, Oceania, South Asia, Western and Central Europe and Western Asia; the regional classification is adapted from IFASTAT. The group of international suppliers ($m = 5$) includes Ma'aden, Mosaic, OCP, combined PhosAgro and EuroChem companies, and combined China's export companies. We run computational experiments for DAP/MAP markets using the three-stage procedure. In the first stage, we sample input parameters from the observed data using the bootstrap method. The set of inputs includes exogenous parameters for supply, demand, and production and trade costs. In the second stage, we run the procedure of generalized English auction (as our computational experiment) for all bootstrap replications of input parameters and calculate response outputs of interest for each computed equilibrium. The auction runs are based on sampling of inputs, and the response outputs can be viewed as observations obtained in a statistical experiment. In the third stage, we assess variability of experimental outputs.

*3.2 Data processing*

We compile a dataset with DAP/MAP trade flows in the integrated market for the period from 2013 to 2017 based on the annual DAP and MAP international trade flows (UN Comtrade 2020); a trade flow from a supplier to a region is estimated as a sum of values, for which import country belongs to the region in question and the trade partner is a country of residence for the supplier (Russia and Lithuania are attributed as a country of residence for PhosAgro/EuroChem). In case of Mosaic and PhosAgro/EuroChem, domestic supply to a country of residence (estimated from company's data and UN Comtrade export flows), is added to corresponding regional trade flow. Our dataset is harmonized with IFASTAT DAP/MAP apparent consumption by region (IFA 2020a); supply of domestic industry in a region is estimated as the difference between apparent consumption and the sum of the calculated trade flows from international suppliers. All volumes are converted to $P_2O_5$ units with conversion factors: 0.46 (DAP), 0.52 (MAP), 0.44 (MAP China) and 0.49 (DAP/MAP).

Phosphorus application rates by crop and country are calculated from the crop production and phosphorus fertilizer use data. The country-level crop production data is available from FAOSTAT (FAO 2020). The phosphorus fertilizer use data by crop and country is derived from the IFASTAT national-level assessment of 2014 phosphorus fertilizer use by crop for 27 countries, the EU, and the rest of the world (IFA-IPNI 2017). Original data on fertilizer use is harmonized to replicate IFASTAT country- and region-level $P_2O_5$ fertilizer consumption (IFA 2020b). The EU fertilizer use data is disaggregated by country in proportions to FAOSTAT country-level crop production data for 2014. The crop-level data for the rest of the world is disaggregated by region in proportions to region-level FAOSTAT crop production data. If there is no country data on phosphorus fertilizer use for production of some crop in 2014, then the minimum application rate attributed to country's region and crop in question is used instead.

*3.3 Demand*

This paper covers FAO alternative future scenarios for world agriculture concerning global intensive farming development by year 2030. The first scenario, "Business as Usual" (BAU), describes development along current trends and policies with outstanding challenges for food and agricultural systems remaining unaddressed. The second scenario, "Stratified Societies" (SSS), outlines a future with exacerbated socio-economic and technological inequalities across countries and throughout different layers of societies. In this scenario both equity and sustainable production are more seriously challenged than under the "Business as Usual" scenario (FAO 2018). We shall make two standing assumptions on the demand-side of the model: 1) phosphorus fertilizer demand changes with the amount of crop produced, and does not change as much with the price according to the build-up and maintenance approach to soil nutrient management (Reetz 2016); 2) the input-output relationship between phosphorus fertilizer use and crop production observed in 2014 holds over time (FAO 2000). Under these assumptions, the phosphorus fertilizer use is derived from crop- and country-level data on phosphorus application rates and scenario crop production volumes in 2030; the scenario estimates are then aggregated by region.

We put forward a two-stage linear regression model for DAP/MAP demand in region $j$:

$$y^j = \beta x^j + u_1^j,$$
$$x^j = \alpha z^j + u_2^j,$$
$$u_{11}^j, \dots, u_{1p}^j \sim F_1(0, \sigma_1^2), \qquad u_{21}^j, \dots, u_{2p}^j \sim F_2(0, \sigma_2^2).$$

Here, $y^j$ is a $p \times 1$ data vector of DAP/MAP demand, $x^j$ is a $p \times 1$ data vector of phosphorus fertilizer consumption (applications to crops, pastures, forests, fish ponds, turf, ornamentals) and $z^j$ is a $p \times 1$ data vector of phosphorus fertilizer use in application to crops. The terms $u_1^j$ and $u_2^j$ are $p \times 1$ vectors of independent and identically distributed errors. Observed data is collected for years 2007-2017 from IFASTAT DAP/MAP apparent consumption ($y^j$), IFASTAT $P_2O_5$ fertilizer consumption ($x^j$) and region-level phosphorus fertilizer use derived from FAOSTAT country-level crop production volumes ($z^j$). Original values are smoothed using central moving average with span 3.

The bootstrap data-generating process for scenario DAP/MAP demand $d_j$ replicates two-step procedure. First, simulated data is calculated for observed $z^j$ and $\alpha$, $\beta$, $u_1^j$ and $u_2^j$ estimated from the data using two-stage least squares with wild unrestricted residual bootstrap (Davidson and MacKinnon 2010). Secondly, bootstrap replication of $d_j$ is computed as a predicted value for scenario value of $z^j$ and parameter estimates obtained as in the first step but using the simulated data instead of the observed one.

### 3.4 Supply

The observed data on producer's supply is composed from the 2013-2017 DAP/MAP trade flows data aggregated by supplier. We adopt assumption from the IFA "Commodity Classic" scenario that large-scale suppliers will strive to maintain their market shares in global DAP/MAP demand by adding new production capacities as the market grows; mean of the observed data scaled proportionally to the global demand is taken as an original estimate of supplier's market share. We compose a joint sample of observed deviations from the mean across suppliers. The bootstrap replications of $s_i$ are then derived from bootstrap scenario values of global demand and original estimate with disturbance term resampled from the joint sample using wild bootstrap.

### 3.5 Trade and production costs

We assume that trade costs shift in inverse proportion to the growth of a market share in the global demand as market becomes more attractive to a supplier. To this end, we apply a regression model: $w = \gamma v + \epsilon$, $\epsilon_1, \ldots, \epsilon_r \sim F(0, \sigma^2)$, where $w$ is a $r \times 1$ data vector of year-over-year change in the trade cost, $v$ is a $r \times 1$ data vector of year-over-year change in the market share, and $\epsilon$ is a $r \times 1$ vector of independent and identically distributed errors. Annual data on DAP/MAP trade flows gives observations on flows associated with a single model equilibrium.

Due to economies of scale, the cost of production for local suppliers, $c_{0j}$, is required to be inversely proportional to the share of market $j$ in global demand. From observed trade flows and local supply, we can derive at first equilibrium market prices, and then, trade costs measured relative to a reference market; the values of $w$ are represented by a joint sample of changes in the relative trade costs measured relative to a reference year. Nominal values of market shares in the global demand are adjusted for the growth of a market share of the reference market measured relative to the reference year; the values of $v$ are taken as changes in these observed real values measured relative to the reference year.

The bootstrap data-generating process for scenario trade costs $t_{ij}$ replicates one-step procedure. Bootstrap replication of $t_{ij}$ is derived from predicted change in the trade cost computed for scenario change of the market share in the global demand and regression parameters $\gamma$ and $\epsilon$ estimated from the data using ordinary least squares with wild unrestricted residual bootstrap. If there is no supplier-region trade flow in 2013-2017 data, then trade between supplier and region in question is not allowed in the model. Bootstrap replications of $a$ and $c_{0j}$ are computed so that supply of domestic industry on each market equates bootstrap scenario demand at unit market price.

**4 Results**

*4.1 Spatial patterns in phosphorus demand*

The phosphorus fertilizer use in application to crops is projected to increase globally by 20.3% in BAU and by 23.5% in SSS (Table 1). The changes in the amount of fertilizer applied vary widely throughout the regions. Global growth is driven by East Asia which accounts for 40.4% of the growth in BAU and for 41% of the growth in SSS. North America's contribution to global growth increases from 5.7% in BAU to 11.5% in SSS; contrariwise, South Asia's share reduces from 26.1% in BAU to 19.2% in SSS. The values for the rest of the regions vary insignificantly across scenarios; Latin America accounts for close to 9% of global growth, Africa and Western and Central Europe – both close to 5-6%, Western Asia – close to 4%, Eastern Europe and Central Asia – close to 2%, Oceania – close to 1%.

| Region | Data | Business as Usual | Stratified Societies |
|---|---|---|---|
| Africa | 1.41 | 1.92 | 1.89 |
| East Asia | 16.90 | 20.34 | 20.94 |
| Eastern Europe and Central Asia | 1.13 | 1.29 | 1.35 |
| Latin America | 7.05 | 7.84 | 7.99 |
| North America | 4.88 | 5.36 | 6.00 |
| Oceania | 0.61 | 0.70 | 0.75 |
| South Asia | 7.12 | 9.34 | 9.01 |
| Western and Central Europe | 1.93 | 2.35 | 2.56 |
| Western Asia | 0.87 | 1.25 | 1.23 |
| World | 41.89 | 50.40 | 51.73 |

**Table 1.** Phosphorus fertilizer use in application to crops (Mt) derived for 2015-2017 data (central moving average with span 3) and projected for 2030 under FAO global crops production scenarios (FAO 2018).

In bootstrap experiments, DAP/MAP demand is predicted to increase globally on average by 16% in BAU and by 18.9% in SSS (Table 2). In both scenarios, Latin America contributes on average close to 3% of global growth in DAP/MAP less than in the case of all phosphate products, which can be explained by a widespread use of single and triple superphosphates in the region. SSS scenario is characterized by the growth of agricultural production in developed countries and by the drop in crops production in developing countries compared to BAU. Growth of North America's DAP/MAP regional demand changes on average from 15% in BAU to 29% in SSS shifting region's contribution to global growth from 12% to 19%. Western and Central Europe growth changes from 8% in BAU to 17% in SSS, which corresponds to 2% and 4% in global growth. Contrariwise, growth of South Asia's DAP/MAP demand decreases on average from 26% in BAU to 21% in SSS, equivalent to 30% and 20% of global growth. The changes in other regions alter insignificantly regions' contributions to global growth across scenarios. East Asia shows 18-21% average regional growth in DAP/MAP demand (45% of global growth), Africa - close to 14% (3% of global growth), Eastern Europe and Central Asia – 15-20% (3% of global growth), Western Asia – 8% (1% of global growth), Oceania – 3-11% (1% of global growth).

| Region | Data | Business as Usual | | Stratified Societies | |
|---|---|---|---|---|---|
| | | Mean | SD | Mean | SD |
| Africa | 1.10 | 1.28 | 0.16 | 1.25 | 0.15 |
| East Asia | 12.39 | 14.66 | 1.23 | 15.05 | 1.26 |
| Eastern Europe and Central Asia | 0.87 | 1.00 | 0.05 | 1.05 | 0.05 |
| Latin America | 4.38 | 4.52 | 0.47 | 4.59 | 0.46 |
| North America | 3.90 | 4.51 | 0.38 | 5.04 | 0.38 |
| Oceania | 0.87 | 0.90 | 0.13 | 0.97 | 0.13 |
| South Asia | 5.83 | 7.33 | 0.78 | 7.04 | 0.80 |
| Western and Central Europe | 1.45 | 1.57 | 0.16 | 1.70 | 0.17 |
| Western Asia | 0.64 | 0.69 | 0.10 | 0.69 | 0.10 |

**Table 2.** DAP/MAP demand (Mt) observed in 2015-2017 (central moving average with span 3) and predicted for 2030 under FAO global crops production scenarios (FAO 2018). Number of bootstrap samples 1000.

*4.2 Density of the global market*

Simulated market shares for China's export, Ma'aden, Mosaic, OCP and PhosAgro/EuroChem were 10.1%, 3.4%, 11.5%, 5% and 7.3% respectively, with standard deviations of 1% (here and below number of bootstrap replications is 1000). The number of trade connections differs across regions. There are 4 international trading partners with North America and Western Asia in the trade network, 3 - with Western and Central Europe and 1 (PhosAgro/EuroChem) with Eastern Europe and Central Asia; the rest of the regions can trade with all suppliers. South Asia is the most attractive market for international suppliers with average market price of supplier's entry at cost value (equivalent to trade costs in the model) standing at 0.01 in BAU and 0.02 in SSS. Latin America, North America and East Asia have moderate average floor prices of entry: 0.04, 0.06, 0.07 in BAU, and 0.04, 0.05, 0.06 in SSS, respectively. The floor entry prices increase for other regions; the average values for Eastern Europe and Central Asia, Western and Central Europe, Oceania, Africa and Western Asia are 0.12, 0.12, 0.14, 0.15, 0.16 in BAU, and 0.11, 0.11, 0.14, 0.15, 0.15 in SSS, respectively.

Computational experiments point out diverse changes to the amount of competition on regional markets (Table 3). The group of the least attractive markets to international suppliers (by market price of supplier's entry at cost value) sees growth in market concentration. The Herfindahl–Hirschman index increases by 0.26, 0.23, 0.42, 0.25 and 0.29 in BAU and by 0.25, 0.21, 0.39, 0.26 and 0.30 in SSS from the observed averages of 0.49, 0.23, 0.19, 0.4 and 0.5 for Eastern Europe and Central Asia, Western and Central Europe, Oceania, Africa and Western Asia respectively. The structure of competition does not deviate from the observed averages of 0.08, 0.25 and 0.28 in Latin America, North America, and South Asia. Contrariwise, East Asia's index will decrease by 0.14 in BAU and 0.15 in SSS from the high average concentration of 0.82 (this regional market is highly concentrated due to China's self-sufficiency in DAP/MAP). The increase in competition is grounded in the growth of international supply to the region. The share of domestic supply in East Asia's demand reduces by 7% in BAU and by 8% in SSS from the observed average of 92% (see Table 4). The expansion of domestic industry accounts for increased concentration in the regions of Eastern Europe and Central Asia, and Western and Central Europe, where local supplies grow by 33% and 26% in BAU and by 34% and 26% in SSS from the observed average shares in regional demand of 40% and 27% respectively. Contrariwise, there is no effect on market concentration in North America, where market share of local suppliers grows by 8% in both scenarios from the observed average share of 38%.

| Region | 2013 | 2014 | 2015 | 2016 | 2017 | Business as Usual | | Stratified Societies | |
|---|---|---|---|---|---|---|---|---|---|
| | | | | | | Mean | SD | Mean | SD |
| Africa | 0.35 | 0.40 | 0.46 | 0.40 | 0.40 | 0.66 | 0.25 | 0.67 | 0.25 |
| East Asia | 0.79 | 0.81 | 0.82 | 0.85 | 0.82 | 0.68 | 0.13 | 0.67 | 0.12 |
| Eastern Europe and Central Asia | 0.47 | 0.48 | 0.40 | 0.52 | 0.56 | 0.74 | 0.24 | 0.74 | 0.24 |
| Latin America | 0.10 | 0.06 | 0.13 | 0.08 | 0.04 | 0.12 | 0.05 | 0.12 | 0.05 |
| North America | 0.31 | 0.26 | 0.25 | 0.23 | 0.19 | 0.29 | 0.08 | 0.27 | 0.07 |
| Oceania | 0.24 | 0.15 | 0.19 | 0.19 | 0.16 | 0.61 | 0.26 | 0.58 | 0.26 |
| South Asia | 0.25 | 0.23 | 0.17 | 0.26 | 0.27 | 0.20 | 0.06 | 0.21 | 0.06 |
| Western and Central Europe | 0.24 | 0.20 | 0.28 | 0.24 | 0.20 | 0.46 | 0.20 | 0.45 | 0.19 |
| Western Asia | 0.60 | 0.47 | 0.51 | 0.49 | 0.43 | 0.79 | 0.25 | 0.80 | 0.25 |

**Table 3.** Market concentration observed and predicted for 2030 under FAO global crops production scenarios (FAO 2018). Number of bootstrap samples 1000.

| Region | 2013 | 2014 | 2015 | 2016 | 2017 | Business as Usual | | Stratified Societies | |
|---|---|---|---|---|---|---|---|---|---|
| | | | | | | Mean | SD | Mean | SD |
| Africa | 0.65 | 0.69 | 0.73 | 0.68 | 0.68 | 0.72 | 0.27 | 0.72 | 0.27 |
| East Asia | 0.91 | 0.91 | 0.92 | 0.93 | 0.92 | 0.85 | 0.07 | 0.84 | 0.07 |
| Eastern Europe and Central Asia | 0.67 | 0.31 | 0.51 | 0.27 | 0.24 | 0.73 | 0.30 | 0.74 | 0.29 |
| Latin America | 0.30 | 0.29 | 0.42 | 0.37 | 0.31 | 0.36 | 0.06 | 0.36 | 0.06 |
| North America | 0.37 | 0.44 | 0.39 | 0.36 | 0.37 | 0.47 | 0.09 | 0.46 | 0.07 |
| Oceania | 0.49 | 0.44 | 0.45 | 0.16 | 0.37 | 0.40 | 0.34 | 0.42 | 0.33 |
| South Asia | 0.54 | 0.50 | 0.39 | 0.54 | 0.55 | 0.47 | 0.06 | 0.48 | 0.06 |
| Western and Central Europe | 0.31 | 0.30 | 0.20 | 0.26 | 0.29 | 0.53 | 0.25 | 0.53 | 0.23 |
| Western Asia | 0.81 | 0.73 | 0.75 | 0.74 | 0.70 | 0.72 | 0.35 | 0.72 | 0.35 |

**Table 4.** Market share of local suppliers observed and predicted for 2030 under FAO global crops production scenarios (FAO 2018). Number of bootstrap samples 1000.

On the supply side of the trade network, the sales of international suppliers become less differentiated across regions. The index of diversification (Table 5) drops sharply by 0.22 and 0.16 in BAU and by 0.2 and 0.16 in SSS from the observed averages of 0.56 and 0.83 for Ma'aden and OCP companies, respectively. China's export and PhosAgro/EuroChem experience less change with diversification decrease by 0.07 (for both) in BAU and 0.05 and 0.07 in SSS from the observed averages of 0.74 and 0.86 respectively. Diversification of Mosaic sales vary insignificantly in both scenarios compared to the past data.

| Supplier | 2013 | 2014 | 2015 | 2016 | 2017 | Business as Usual | | Stratified Societies | |
|---|---|---|---|---|---|---|---|---|---|
| | | | | | | Mean | SD | Mean | SD |
| China's export | 0.69 | 0.81 | 0.68 | 0.73 | 0.78 | 0.68 | 0.15 | 0.69 | 0.14 |
| Ma'aden | 0.57 | 0.52 | 0.40 | 0.61 | 0.70 | 0.34 | 0.27 | 0.36 | 0.27 |
| Mosaic | 0.72 | 0.74 | 0.71 | 0.69 | 0.71 | 0.73 | 0.08 | 0.72 | 0.09 |
| OCP | 0.76 | 0.80 | 0.88 | 0.87 | 0.83 | 0.66 | 0.17 | 0.67 | 0.17 |
| PhosAgro/EuroChem | 0.84 | 0.84 | 0.89 | 0.87 | 0.87 | 0.79 | 0.12 | 0.79 | 0.12 |

**Table 5.** Diversification of international suppliers observed and predicted for 2030 under FAO global crops production scenarios (FAO 2018). Number of bootstrap samples 1000.

## 5 Policy implications

Phosphate fertilizer is an essential non-substitutable input into industrial agriculture. The future level of agricultural intensification and the type of crops produced will determine the growth of phosphorus demand across regions. Our projections of DAP/MAP fertilizer consumption (for direct application and for blending) suggest that current spatial distribution of global demand will be strengthened by 2030 with two biggest markets, East Asia and South Asia, contributing the most to the global growth. Additionally, we expect stronger DAP/MAP demand growth for North America and Western and Central Europe, and weaker demand growth for South Asia in the scenario of sharpened socio-economic and technological inequalities across countries, than in the scenario of agriculture systems developing along current trends.

The adaption of global flexible sales strategy by international suppliers has bound the regional markets to one another due to phosphate industry consolidation and high entry barrier into the international market. Simulations of matching market equilibrium for selected input parameters generate similar structures of DAP/MAP supply network for both scenarios of intensive farming evolution by 2030. We anticipate reduction in market density in response to the non-uniform DAP/MAP demand growth across regions. Market concentration will increase for small-scale markets as their demand grows slower than in major DAP/MAP consumption regions. Competition will increase in East Asia, while remaining stable in other large-scale regions comprising South Asia, North America, and Latin America. On the supply side, simulated equilibria indicate that large-scale cost-efficient multi-market suppliers will concentrate on fewer markets relative to their current regional sales distribution. In this situation, the less dense structure of the distributed global market becomes more vulnerable to disruptions in major trade flows as the price adjustment process must compensate for larger volume substitution than before. The regions of Eastern Europe and Central Asia, Western and Central Europe and, to a lesser extent, North America demonstrate a high rate of import substitution by local suppliers within regional market vicinity by the end of the next decade. This rate of substitution will require additional capital investment in the domestic industry; otherwise, the abrupt increase in demand can lead to price growth if local suppliers fail to introduce new capacities.

## 6 Conclusion

An economic model of many-to-many matching market for annual DAP/MAP commodity trade has been presented and applied here to assess the impact of phosphate fertilizer industry consolidation along with the emerged trade structure, on future phosphorus supply for region-level crops production in the FAO scenarios for world agriculture by year 2030. The findings presented in this paper should be considered within the limits of the model and adopted assumptions. Due to the restricted public access to the company-level data, the model was applied on the regional basis with additional assumptions being made for supply, and production and trade costs estimation. Availability of spatially explicit expert-based projections on phosphorus fertilizer use in agriculture represents another challenge; this paper adopted statistical approach to estimation of future DAP/MAP demand with additional assumptions being made based on the information about phosphorus fertilizer field application. The bootstrap method was used to propagate uncertainty in the values of input parameters into the simulation model and assess then variability of experimental outputs.

The model of two-sided matching market accounts for the structure of the spatially distributed global commodity market. It therefore yields a more accurate representation of real markets, integrating the supply-demand dynamics of individual markets into the single mechanism. This property makes the model suitable for future projections for phosphorus commodity fertilizer trade, where a small group of large-scale companies dominates the international market distributing phosphate products across all world regions. The application of this modeling approach to 2030 FAO global crops production scenarios can serve as an example of the impact of market interdependence and consolidated supply on the density of DAP/MAP supply network for regional agricultural production, and as a result, on its structural stability for possible supply disruptions.

1 **Appendix**

2 **Table A.1.** Trade costs simulated for 2030 under FAO Business as Usual scenario (FAO 2018). Number of bootstrap samples 1000.

| Region | China's export | | Ma'aden | | Mosaic | | OCP | | PhosAgro/EuroChem | |
|---|---|---|---|---|---|---|---|---|---|---|
| | Mean | SD | Mean | SD | Mean | SD | Mean | SD | Mean | SD |
| Africa | 0.17 | 0.04 | 0.15 | 0.05 | 0.16 | 0.04 | 0.13 | 0.04 | 0.14 | 0.04 |
| East Asia | 0.07 | 0.05 | 0.08 | 0.05 | 0.07 | 0.05 | 0.06 | 0.04 | 0.06 | 0.05 |
| Eastern Europe and Central Asia | - | - | - | - | - | - | - | - | 0.12 | 0.04 |
| Latin America | 0.07 | 0.04 | 0.06 | 0.04 | 0.03 | 0.03 | 0.03 | 0.03 | 0.03 | 0.03 |
| North America | 0.10 | 0.04 | - | - | 0.03 | 0.03 | 0.06 | 0.04 | 0.06 | 0.04 |
| Oceania | 0.17 | 0.04 | 0.15 | 0.05 | 0.15 | 0.04 | 0.13 | 0.04 | 0.11 | 0.04 |
| South Asia | 0.00 | 0.01 | 0.00 | 0.01 | 0.02 | 0.03 | 0.02 | 0.03 | 0.01 | 0.02 |
| Western and Central Europe | 0.15 | 0.04 | - | - | - | - | 0.10 | 0.04 | 0.10 | 0.04 |
| Western Asia | 0.18 | 0.05 | 0.16 | 0.05 | - | - | 0.14 | 0.04 | 0.15 | 0.04 |

Table A.2. Trade costs simulated for 2030 under FAO Stratified Societies scenario (FAO 2018). Number of bootstrap samples 1000.

| Region | China's export | | Ma'aden | | Mosaic | | OCP | | PhosAgro/EuroChem | |
|---|---|---|---|---|---|---|---|---|---|---|
| | Mean | SD | Mean | SD | Mean | SD | Mean | SD | Mean | SD |
| Africa | 0.17 | 0.04 | 0.14 | 0.04 | 0.16 | 0.04 | 0.13 | 0.04 | 0.14 | 0.04 |
| East Asia | 0.06 | 0.05 | 0.07 | 0.05 | 0.07 | 0.05 | 0.05 | 0.04 | 0.05 | 0.04 |
| Eastern Europe and Central Asia | - | - | - | - | - | - | - | - | 0.11 | 0.04 |
| Latin America | 0.06 | 0.04 | 0.05 | 0.04 | 0.03 | 0.03 | 0.02 | 0.03 | 0.02 | 0.03 |
| North America | 0.08 | 0.04 | - | - | 0.02 | 0.03 | 0.04 | 0.03 | 0.05 | 0.03 |
| Oceania | 0.16 | 0.04 | 0.14 | 0.04 | 0.15 | 0.04 | 0.13 | 0.04 | 0.10 | 0.04 |
| South Asia | 0.01 | 0.02 | 0.01 | 0.02 | 0.03 | 0.03 | 0.03 | 0.03 | 0.02 | 0.03 |
| Western and Central Europe | 0.14 | 0.04 | - | - | - | - | 0.10 | 0.04 | 0.10 | 0.04 |
| Western Asia | 0.18 | 0.04 | 0.15 | 0.04 | - | - | 0.14 | 0.04 | 0.14 | 0.04 |

**Code availability**

The code for the special case of generalized English auction developed in this study is available via a public repository (Shchiptsova 2021). The code for data processing and statistical routines developed in this study is available via a public repository (Shchiptsova 2022).

[1] Shchiptsova, A., 2021. A special case of generalized English auction: commodities-auction: v1.0.1. Zenodo.

https://doi.org/10.5281/zenodo.5511774

[2] Shchiptsova, A., 2022. Routines for the case of DAP/MAP matching model. Zenodo. https://doi.org/10.5281/zenodo.6457219